\documentclass[sigconf, nonacm]{acmart}

\usepackage[utf8]{inputenc} 
\usepackage[T1]{fontenc}    
\usepackage{hyperref}       
\usepackage{url}            
\usepackage{booktabs}       
\usepackage{amsfonts}       
\usepackage{nicefrac}       
\usepackage{microtype}      
\usepackage{xcolor}         
\usepackage{enumitem}
\usepackage{xspace}
\usepackage{graphicx}
\usepackage{bm}
\usepackage{amsmath}
\usepackage{mathtools}
\usepackage{wrapfig}  
\usepackage{multirow}
\usepackage{caption}
\usepackage[table]{xcolor}
\usepackage{float}
\usepackage{tcolorbox} 
\usepackage{varwidth} 
\tcbuselibrary{breakable} 
\tcbuselibrary{skins} 
\usepackage[dvipsnames,svgnames]{xcolor} 
\colorlet{MorandiGray}{Gray!40!White}     

\definecolor{mypink}{HTML}{ff8383}
\definecolor{myorange}{HTML}{fed06e}
\definecolor{myblue}{HTML}{74bcd5}
\definecolor{mygreen}{HTML}{0a8a8b}
\definecolor{myyellow}{HTML}{fed06e}
\definecolor{mygrey}{HTML}{f5f6f4}
\AtBeginDocument{%
  }

\newcommand{\ga}{{Gaze Archive}\xspace}
\newcommand{\mtd}{\textsc{GaHMA}\xspace}
\newcommand{\mtdns}{\textsc{GaHMA}}
\newcommand{\bmk}{\textsc{GaVER}\xspace}
\newcommand{\mtdf}{\textsc{Gaze-aware Hierarchical Memory Archiving}\xspace}
\newcommand{\bmkf}{\textsc{Gaze-annotated Visual Encoding and Retrieval}\xspace}

\newcommand{\lm}{{LVLMs}\xspace}

\begin{document}

\title{Gaze Archive: Enhancing Human Memory through Active Visual Logging on Smart Glasses}

\author{Haoxin Ren}
\affiliation{
  \institution{State Key Lab. of VR Technology and Systems \\ Beihang University}
  \city{Beijing}
  \country{China}
}
\email{19231135@buaa.edu.cn}

\author{Feng Lu}
\authornote{Corresponding author}
\affiliation{
  \institution{State Key Lab. of VR Technology and Systems \\ Beihang University}
  \city{Beijing}
  \country{China}
}
\email{lufeng@buaa.edu.cn}

\renewcommand{\shortauthors}{Ren et al.}

\begin{abstract}
  People today are overwhelmed by massive amounts of information, leading to cognitive overload and memory burden. Traditional visual memory augmentation methods are either effortful and disruptive or fail to align with user intent. To address these limitations, we propose \ga, a novel visual memory enhancement paradigm through active logging on smart glasses. It leverages human gaze as a natural attention indicator, enabling both intent-precise capture and effortless-and-unobtrusive interaction. To implement \ga, we develop \mtd, a technical framework that enables compact yet intent-aligned memory encoding and intuitive memory recall based on natural language queries. Quantitative experiments on our newly constructed \bmk dataset show that \mtd achieves more intent-precise logging than non-gaze baselines. Through extensive user studies in both laboratory and real-world scenarios, we compare \ga with other existing memory augmentation methods. Results demonstrate its advantages in perceived effortlessness, unobtrusiveness and overall preference, showing strong potential for real-world deployment.
\end{abstract}


\begin{teaserfigure}
  \includegraphics[width=\textwidth]{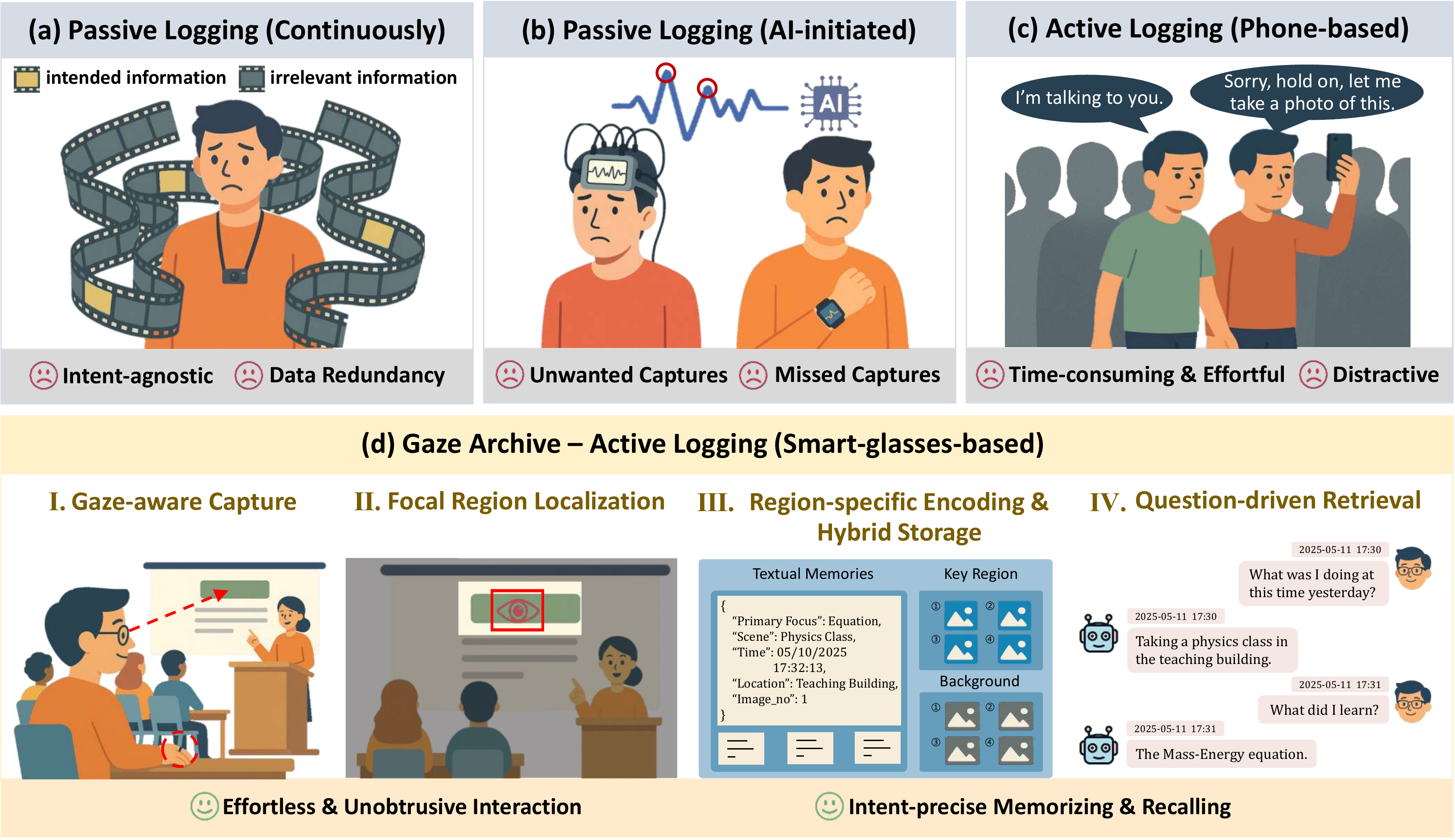}
  \caption{\textbf{Illustration of existing visual memory augmentation methods and \ga.} (Upper) The limitations of existing methods: (a) continuous logging leads to data redundancy; (b) AI-initiated logging is not accurate enough in identifying user intent; (c) phone-based active logging is effortful and distractive. (Lower) (d) \ga, a gaze-driven paradigm for intent-precise and effortless-and-unobtrusive visual memory augmentation through active logging on smart glasses. Inspired by human vision, our approach uses eye gaze to identify focal content for efficient encoding and intuitive recall through natural language questions.}
  \Description{Enjoying the baseball game from the third-base
  seats. Ichiro Suzuki preparing to bat.}
  \label{fig:teaser}
\end{teaserfigure}


\maketitle

\section{Introduction}
Memory, as a crucial component of human cognitive ability, plays an indispensable role in individual learning, working and daily life \cite{gerard1953memory}. However, with the increasing complexity of human activities, people today are exposed to more information than ever, leading to overload that harms attention and memory \cite{arnold2023dealing}. Psychological studies show that individuals struggle to focus and recall under such cognitive burden \cite{langerock2025cognitive}. As a result, people today tend to rely on external tools like search engines rather than memorize on their own, a phenomenon known as the “Google Effect” \cite{sparrow2011google}.

To address this, many HCI researchers have explored \textbf{visual logging} as a way of externalizing parts of personal memory into digital archives that can be searched and revisited later. Within this space, existing approaches can be generally categorized into two paradigms: \textbf{passive logging} and \textbf{active logging}. Lifelogging is one representative form of passive logging, in which wearable cameras continuously record user's daily life without manual intervention \cite{harvey2016remembering}. While creating full visual archives, it also generates vast amounts of redundant and irrelevant data, thereby leading to a cluttered memory collection and undermining retrieval efficiency, as shown in Fig.~\ref{fig:teaser}(a).

To reduce redundancy, AI-initiated passive logging methods based on physiological sensing have been proposed. For example, electroencephalography (EEG) \cite{jiang2019memento}, galvanic skin response (GSR) \cite{chan2020biosignal}, and other biosignals have been used to detect moments of heightened emotional arousal, under the assumption that emotionally salient events are more likely to be worth remembering. In these systems, fluctuations in physiological signals automatically trigger the initiation and termination of recording, reducing the capture of irrelevant footage. While this approach represents a step toward intent-aligned memory capture, it comes with its own limitations. As illustrated in Fig.~\ref{fig:teaser}(b), since the reliability of emotion detection algorithms remains imperfect, errors in identifying significant moments may result in the omission of intended information or the retention of irrelevant content \cite{ksibi2021overview,cai2024pandalens}.

Another line of work has explored user-initiated logging via smartphones \cite{li2025omniquery}. In such systems, image capture is actively triggered by user, ensuring that the recorded material is intentionally selected. While this method offers greater alignment with user's intent, it introduces practical drawbacks, as shown in Fig.~\ref{fig:teaser}(c). Manually retrieving a phone, launching the camera, and framing a shot can be time-consuming, effortful and distracting, especially when user is engaged in another primary task. Moreover, such overt capture actions may draw unwanted attention in public settings, resulting in a suboptimal user experience in social scenarios.

Thus, we distill two fundamental requirements for effective visual memory enhancement systems.
(1) \textbf{Intent-precise}: recorded content should match user intent, which is unpredictably influenced by individual factors; 
(2) \textbf{Effortless-and-unobtrusive}: interaction should minimize physical/cognitive load while remaining socially unnoticeable. 
No current solution fully achieves both.

To address these dual requirements, we propose \ga, a new paradigm for visual memory enhancement that leverages gaze-based active logging on AI-powered smart glasses to enable both intent-precise and effortless-and-unobtrusive memory assistance, as depicted in Fig.~\ref{fig:teaser}(d). Our key insight is that human gaze naturally reflects attention — a core mechanism in biological vision and a well-established cognitive indicator. By leveraging gaze focusing as a natural guiding cue in combination with AI's semantic interpretation capabilities, we greatly reduce the amount of visual data to encode and store, as naturally performed by the brain. In terms of hardware form factor, \ga uses subtle ring-based operation to trigger active logging on smart glasses, enabling both effortless and unobtrusive interaction. This mechanism of identifying and realizing user intent through smart device-based human-AI collaboration, exemplifies the characteristics of emerging AI paradigms such as cobodied/symbodied AI \cite{lu2025towards}.

Building upon the \ga paradigm, we develop \mtdf (\mtd), a technical framework for gaze-guided visual memory enhancement. It integrates four core components: (1) smart glasses for unobtrusive gaze and scene capture; (2) adaptive scene partitioning to identify focal regions based on foveal vision and semantic context; (3) region-specific hierarchical encoding to generate compact yet informative memory; and (4) question-driven retrieval for intuitive memory access through natural language queries.
Together, these components realize a seamless loop of visual memory acquisition, encoding, and recall.

To evaluate methods under this new paradigm, we conducted a two-stage investigation. In the first stage, we constructed \bmkf (\bmk), a benchmark dataset with gaze-annotated real-world scenes and associated question-answer (QA) pairs, and carried out a large-scale quantitative experiment under varied conditions. Experimental results show that \mtd achieves higher recall accuracy than non-gaze baselines with significantly lower storage cost. In the second stage, we compared \ga with existing visual memory augmentation methods (Phone-based and Lifelogging) through both in-lab and in-situ user studies. The results confirm its advantages in perceived effortlessness, unobtrusiveness, and user preference for daily use.

In summary, our paper makes the following contributions:
\begin{itemize}
\item We propose \ga, a biologically inspired paradigm for visual memory augmentation through active logging on smart glasses. It supports both intent-precise and effortless-and-unobtrusive memory recording.
\item We propose \mtd, a technical framework for implementing \ga, which incorporates four core components to realize a seamless memory enhancement pipeline. 
\item We conduct large-scale quantitative experiments to evaluate \mtd against non-gaze baselines, and compare \ga with existing visual memory augmentation approaches in both laboratory and real-world scenarios, demonstrating its promising performance and user preference in daily use.

\end{itemize}

\section{Related Work}

\subsection{Personal Memory Augmentation via Visual Logging}

Memory augmentation has always been one of the research hotspots in the field of human-computer interaction (HCI). As early as 1945, Vannevar Bush envisioned a microfilm-based machine called Memex \cite{bush1945}, that could be used to store personal documents, photographs and sound recordings, providing functions such as text annotation and full-text search, thereby extending human memory. With the development of sensor and digital storage technology, the vision of Memex has gradually become achievable \cite{gemmell2002mylifebits}, along with the emergence of the concept of lifelogging \cite{dodge2007outlines}.

Among various modalities, visual information has become the most widely used for memory augmentation, due to its richness and intuitiveness \cite{konkle2010conceptual}. Traditional visual lifelogging systems continuously capture first-person perspectives using wearable cameras \cite{kalokyri2022supporting}, forming extensive visual archives but often leading to data redundancy and retrieval challenges. Recent approaches have explored semantic tagging \cite{byrne2010everyday}, event segmentation \cite{wiik2020multimodal,shah2012lifelogging,gupta2018approaches}, and image/video captioning \cite{shen2024encode,le2016impact} for more efficient organization. Retrieval mechanisms include tag-based filtering \cite{byrne2010everyday}, visual similarity matching \cite{doherty2012experiences}, and natural language queries \cite{shen2024encode,tran2021mysceal,tran2024interactive}. Notably, Shen et al. introduced an AR helmet that captures egocentric images and leverages language encoding for memory storage and retrieval \cite{shen2024encode}. However, these methods still suffer from low storage efficiency due to recording excessive amounts of useless data.

To overcome the redundancy inherent in continuous passive logging, researchers leverage AI technology to capture moments of significance automatically. For instance, Memento uses Electroencephalogram (EEG) to detect emotional peaks and trigger memory logging \cite{jiang2019memento}, while Prompto uses physiological signals like electrodermal activity (EDA) and heart-rate variability (HRV) to infer cognitive load \cite{chan2020prompto}. However, current AI algorithms are not yet perfectly accurate, leading to potential misalignment with user intent. Alternatively, OmniQuery enables explicit active visual logging through smartphone photography by user themselves \cite{li2025omniquery}, but this approach is effortful, distractive and socially obtrusive in public environments \cite{tran2022mysceal,makela2021hidden}.

\subsection{Gaze-based Interaction}
As a natural indicator of visual attention, gaze can be captured using head-mounted displays equipped with eye-trackers. Based on gaze tracking, utilizing gaze for natural interaction has been widely explored in augmented reality (AR) applications. Gaze was mainly used for two scenarios: as sole input signal or in combination with other modalities \cite{wang2025towards}. Gaze-only interaction enables intuitive localization and selection \cite{lee2024snap,plopski2022eye,jacob1990you}, yet it suffers from unconscious mis-touches (the Midas touch problem) \cite{jacob1990you}, where unintentional gaze behaviors are mistakenly interpreted as interaction commands. To address this, various strategies such as dwell time and gaze gestures have been tested for disambiguation \cite{skerjanc1997new,istance2010designing}. However, they may cause fatigue and reduce the interaction efficiency \cite{wang2025towards}. Multimodal extensions combining gaze with speech \cite{li2024omniactions} or gestures \cite{bao2023exploring,wang2020comparing,ren2024eye,shi2023exploring} can create a richer and more expressive experience for user and also better facilitate certain tasks. However, voice commands or overt gestures may raise concerns about public acceptance in social contexts. Wang et al. proposed GazeRing, integrating gaze with subtle finger movements via a Bluetooth ring for discreet interaction \cite{wang2024gazering}. Although approaches like GazeRing show promise for effortless and unobtrusive active visual logging, they have yet to be explored.

Apart from using eye tracking data for explicit input and selection, gaze can also be used to understand human cognitive process and user intention. For instance, GazePointAR \cite{lee2024gazepointar} and G-VOILA \cite{wang2024g} leverage eye gaze to identify areas of interest/referential pointing, thus enabling more accurate interpretations of user intention. In our work, we will further explore the use of gaze as a natural attention indicator to guide intent-precise visual memory augmentation.

\subsection{LLMs/\lm-enhanced Wearable Interactive Systems}
In recent years, wearable devices have attracted sustained attention due to their potential to provide natural and low-disruption interaction experiences \cite{jhajharia2014wearable,neupane2025wearable}. Compared with handheld or stationary devices, wearables enable ever-present and seamlessly accessible interaction with minimal physical and cognitive load, making them particularly suitable for daily assistance. 

Recent advancement in large language models (LLMs) and large vision-language models (\lm) have further enhanced the capabilities of wearable interactive systems \cite{arava2025large}. LLMs endow wearables with natural language understanding and generation capabilities, enabling more intuitive and adaptive access to information and task assistance through conversational interfaces. Moreover, \lm break through the limitations of single text modality, facilitating wearables with visual perception and reasoning capabilities, thus better supporting complex real-world scenarios. For instance, Memoro \cite{zulfikar2024memoro} uses LLMs to realize real-time conversational aid on bone-conduction headsets, while Google Project Astra \cite{project-astra}, GazePointAR \cite{lee2024gazepointar} and AiGet \cite{cai2025aiget} allow users to learn about physical objects and surroundings through natural interactions (e.g., voice or gesture queries) using AR HMDs. These work demonstrate the potential of LLMs/\lm-enhanced wearable systems for more natural and intelligent assistance. In this study, we will further investigate the use of \lm to enable intent-aware visual memory enhancement on smart glasses.

\section{Overview of \ga}

\begin{figure*}
  \centering
  \includegraphics[width=\linewidth]{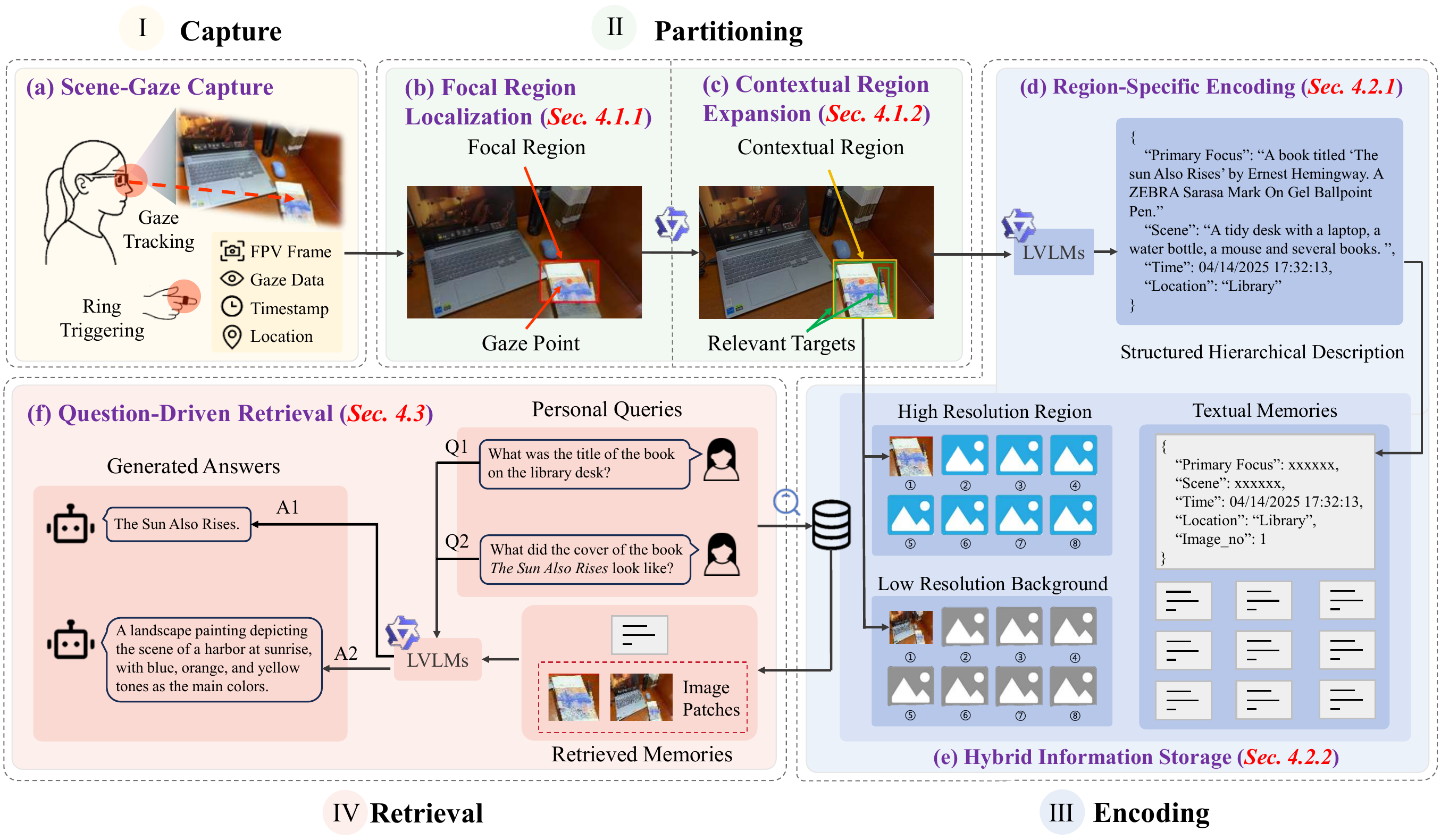}
  \caption{\textbf{Pipeline of \mtd.} \textit{Capture:} (a) smart glasses capture of scene-gaze pairs with other auxiliary information; \textit{Partitioning:} (b) focal region localization based on gaze fixation and foveal vision model; (c) contextual region expansion based on semantic analysis using \lm; \textit{Encoding:} (d) region-specific encoding using \lm; (e) hybrid information storage for better recall quality; \textit{Retrieval:} (f) question-driven retrieval from memory achives and answer with \lm.}
  \label{fig_overview}
\end{figure*}

\subsection{Motivation and Challenges}
Given the limitations of existing visual memory augmentation methods, we propose \ga, a biologically inspired paradigm for gaze-driven visual memory enhancement.
Unlike traditional systems that either passively record everything or rely on intrusive interactions, \ga aims to achieve both:
(1) intent-precise memory capture, and  
(2) effortless-and-unobtrusive interaction.
However, this poses the following technical challenges:

\textbf{Q1: How to capture scene-gaze data rapidly and unobtrusively?}
Manual logging can select target regions but interrupts ongoing tasks and draws attention. The challenge is to collect synchronized scene-gaze pairs during natural visual behavior, with minimal effort and disruption. 

\textbf{Q2: How to identify focal regions that include intended memories?}
Gaze fixations offer only coarse localization. Intended memory often spans a broader area. The challenge is to determine the optimal region to include what matters most while excluding redundancy. 

\textbf{Q3: How to encode memories effectively with limited storage?}
Full-resolution image storage is inefficient; over-summarization leads to recall failure. The challenge is to design a hierarchical strategy that encodes rich information in focal areas and compact cues in the background. 

\subsection{Technical Solution and Pipeline} \label{sec_solution_and_pipeline}

To address the above challenges, we propose \mtdf (\mtd), a technical framework that integrates four core modules to realize \ga paradigm. Consequently, the overall pipeline of \mtd consists of four key stages: capture, partitioning, encoding, and retrieval. Each module is designed to align with natural human visual behavior. They form a seamless loop for gaze-aware visual memory augmentation, as shown in Fig.~\ref{fig_overview}.

\textbf{Stage 1 (Capture)}  
Smart glasses with embedded eye-tracking module records egocentric images, gaze fixations, and auxiliary information synchronously. A lightweight Bluetooth ring enables memory triggers through a simple double-tap when users encounter meaningful content. This interaction is low-effort, socially unobtrusive without disrupting ongoing activities, which addresses \textbf{challenge Q1}.

\textbf{Stage 2 (Partitioning)}  
Given the scene-gaze pairs, the \textbf{focal region partitioning} module (Sec.~\ref{ssec_gahma1}) divides the visual scene into three regions: the {focal region}, the {contextual region} that provides complementary information, and the remaining {peripheral region}. This division leverages human foveal vision modeling and semantic reasoning with \lm, which resolves \textbf{challenge Q2}.

\textbf{Stage 3 (Encoding)}  
Given the partitioned regions, the \textbf{hierarchical memory encoding} module (Sec.~\ref{ssec_gahma2}) uses \lm to generate fine-grained descriptions for the contextual region, while background content is summarized compactly. The results are stored as multimodal entries including structured text, image patches, and embedding vectors. This strategy addresses \textbf{challenge Q3}.

\textbf{Stage 4 (Retrieval)}  
Given the stored memories, the \textbf{question-driven retrieval} module (Sec.~\ref{ssec_gahma3}) supports user queries in natural language about past experiences.  \lm enhance query semantics to resolve ambiguities and search through memory archives. Retrieved entries are then synthesized into natural language responses, completing the visual memory enhancement cycle.

\begin{figure*}
  \centering
      \includegraphics[width=\textwidth]{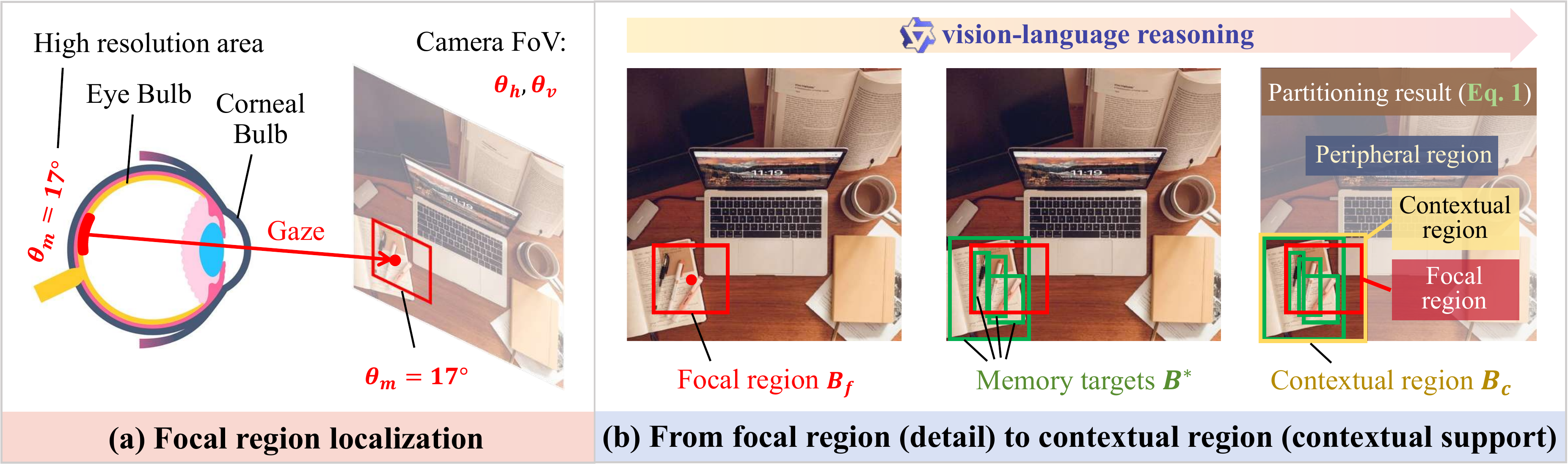}
  \caption{\textbf{Illustration of focal region partitioning.} (a) Focal region is localized based on gaze fixation and foveal vision model; (b) focal region is expanded into contextual region.}
  \label{fig_focalregion}
\end{figure*}

\section{\mtd: \mtdf}\label{sec_gahma}

\subsection{Focal Region Partitioning} \label{ssec_gahma1}
This module divides the scene into three regions: focal (captures intent-aligned content), contextual (provides necessary semantic support), and the remaining peripheral region. 

\subsubsection{Focal region localization}\label{sssec_gahma1}

Let $ I $ denote the input image with resolution $ W \times H $ and $\vec{g} = (x_g, y_g) \in \mathbb{R}^2$ represent the pixel coordinates of the fixation point. Our goal is to define a rectangular region centered at $(x_f, y_f)$ with width $W_f$ and height $H_f$, such that it covers the content users want to memorize.

Based on the biological principle that attention is concentrated around fixations, we assume the focal region center aligns with the gaze fixation, i.e., $(x_f, y_f) = \vec{g}$. 
Regarding the region size, human high-resolution retinal imaging spans a central visual field of approximately $\theta_m \in [10^{\circ}, 20^{\circ}]$, and we set $\theta_m = 17^\circ$ as the effective coverage angle \cite{stra2011peripheral}.
Consequently, the region size are derived by projecting this coverage angle onto the image plane: $W_{f} = W \cdot {\theta_m}/{\theta_{h}}$ and $H_{f} = H \cdot {\theta_m}/{\theta_{v}}$, where $\theta_h$ and $\theta_v$ denote the camera's horizontal and vertical fields of view, as shown in Fig.~\ref{fig_focalregion}(a).
This calculation yields the parameters $\{W_f, H_f, x_f, y_f\}$ for the focal region $ B_f $.

\subsubsection{Contextual region expansion}

Some objects may only be partially captured within the focal region. To address this, we expand it into a broader contextual region that completely encompasses these objects, as shown in Fig.~\ref{fig_focalregion}(b). This enables more accurate memory encoding.

Specifically, we propose an end-to-end vision-language reasoning-based expansion strategy using \lm. The original image annotated with focal region $ B_f $ are fed into \lm. Using structured prompts, it performs joint detection and semantic to infer all the intent-relevant memory targets. 

The output of \lm includes a bounding box set $\{B^*\}$ representing core memory targets.
the contextual region $ B_c$ is computed as the minimal enclosing rectangle over $ B_f $ and all $ B_j \in B^* $, ensuring full coverage of possible intent-relevant memory targets:
\begin{equation}\label{eq_enclose}
    B_c = \text{Enclose}(\{B_j \in B^*\}, B_f).
\end{equation}

\subsection{Hierarchical Memory Encoding}\label{ssec_gahma2}

This module encodes visual memories in a hierarchical manner. The challenge lies in effectively leveraging the focal region and contextual region to balance semantic richness and storage efficiency.

\subsubsection{Region-specific encoding}

We encode visual information into semantic descriptions according to the focal region $ B_f $ and contextual region $ B_c $. This mimics the human visual system's selective processing, where high-acuity perception is concentrated in the fovea, while background information is encoded with lower detail.

Specifically, we apply \lm to generate descriptions $ D_f \in \mathcal{D} $ for the focal region. The input to the \lm includes both $ B_f $ and $ B_c $, together with the full image $I$ and a parameter $ \gamma \in\mathbb{N}$ to control the levels of detail; for the remaining background, we extract a compact summary $ D_b $:
\begin{equation}
    D_f = \text{\lm}(I({B_f}), I({B_c}), p_f, \gamma), \quad D_b = \text{\lm}(I, p_b),
    \label{eq:focal_description}
\end{equation}
where $ p_f $ and $ p_b $ are structured prompts.
Note that $I({B_f})$ and $I({B_c})$ play complementary roles: 
the former captures high-fidelity visual details (``detail encoding''), while the latter establishes contextual grounding to interpret those details meaningfully (``contextual support''). 

The hierarchical strategy captures fine-grained focal details while retaining structural scene content, with minimal additional storage overhead. It balances user-intent fidelity and storage efficiency.

\subsubsection{Hybrid information storage}

After region-specific encoding, we store the generated $ D_f $ and $ D_b $ in a structured format. To balance recall efficiency with storage compactness, we propose a hybrid storage strategy. 

\textbf{Textual entry storage}
We store $ D_f $ and $ D_b $, along with other textual metadata (e.g., timestamp $ t_{stamp} $ and GPS location $ \vec{p}_{gps} $) in the textual memory entry
$    M_t = \{D_f, D_b, t_{stamp}, \vec{p}_{gps}\}$,
which supports QA-based recall tasks with minimal space usage.

\textbf{Image patch storage}  
In addition to textual entries, we optionally retain image patches at different resolutions. High-resolution crops $ I({B_c})$ preserve full semantic context surrounding the gaze, while low-quality versions $ I_{lq} $ of the original image $I$ provide global context with low storage cost. A hybrid memory entry is denoted as $M_{hc} = \{D_f, D_b, t_{stamp}, \vec{p}_{gps}, I(B_c)\}$, $M_{hg} = \{D_f, D_b, t_{stamp}, \vec{p}_{gps}, I_{lq}\}$, or $M_{hcg} = \{D_f, D_b, t_{stamp}, \vec{p}_{gps}, I(B_c), I_{lq}\}$.

\subsection{Question-driven Retrieval}\label{ssec_gahma3}

This module enables users to ask about past experiences in natural language, and generate intent-precise answers based on memory content.


\textbf{Text-only storage setting} A natural language query is fed into \lm for semantic matching to retrieve the relevant memory entry $M_t$, which is also used by \lm to generate fluent responses. 
When the memory archive grows large, all $\{M_t\}$ entries are pre-encoded into embeddings within a RAG framework, allowing fast similarity-based lookup with the query embedding.

\textbf{Hybrid storage setting} A natural language query is first used to retrieve relevant hybrid memory entry $M_{hc}$, $M_{hg}$, or $M_{hcg}$ as in the text-only setting. Then the \lm regenerates the response by using the the combination of textual descriptions and the
associated images ($ I({B_c})$ and/or $ I_{lq} $), which provide richer context for more accurate answers.

\section{Study 1: Large-scale Quantitative Evaluation of \mtd}

\subsection{Research Objectives}
To study the proposed gaze-aware visual memory augmentation system’s performance, we first carry out a large-scale quantitative evaluation, aiming to address the following research questions:

\textbf{RQ1: How does \mtd perform in terms of intent-precise memory encoding?} We deem that an effective visual memory augmentation system should prioritize recording information relevant to user intent while minimizing irrelevant content. This question evaluates whether \mtd achieves intent-aligned encoding compared to non-gaze baselines.

\textbf{RQ2: How do different storage strategies affect recall accuracy and storage efficiency?} Given the inherent trade-off between memory fidelity and storage cost, it is crucial to understand how various storage configurations impact recall performance and resource usage.

\textbf{RQ3: How well does the question-driven retrieval mechanism scale to large memory archives?} A practical memory augmentation system should maintain high retrieval accuracy even as the memory size grows. This question therefore investigates the robustness of the retrieval strategy under varying archive sizes.

\subsection{Dataset: \bmk}
\begin{figure} 
  \centering
  \includegraphics[width=0.3\textwidth]{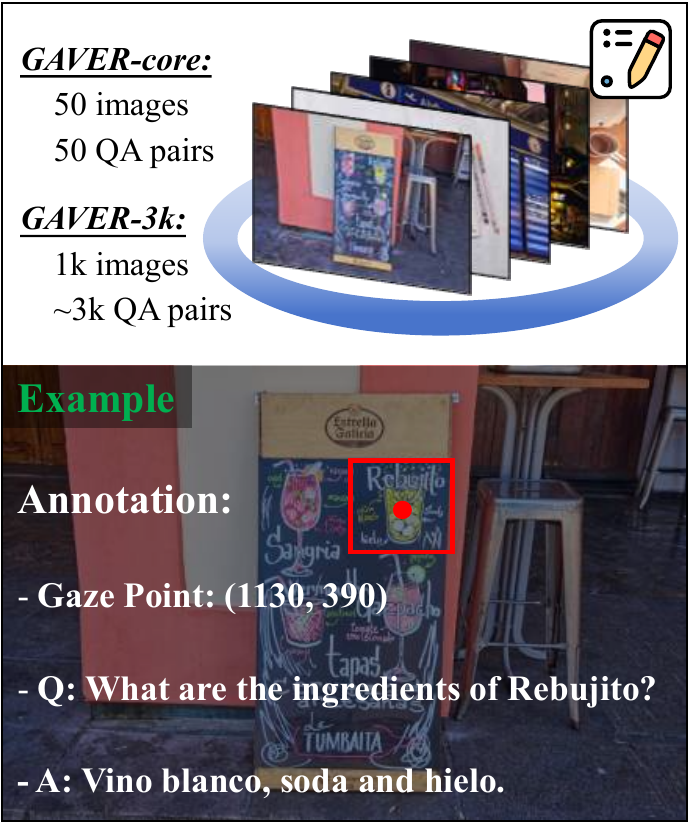}
  \caption{Example image and annotation (gaze point and Q\&A pair) in \bmk dateset.}
  \label{fig_dataset}
\end{figure}

Since no specific dataset supports systematic evaluation of gaze-aware visual memory augmentation systems, we introduce \bmkf (\bmk), a new benchmark dataset with gaze-annotated images and associated question-answer (QA) pairs, as shown in Fig.~\ref{fig_dataset}. \bmk is designed to evaluate key capabilities of intent-precise memory encoding, hierarchical information archiving, and question-guided retrieval. In particular, \bmk contains two subsets.

\textbf{\bmk-core (manually annotated, high-quality subset for in-septh analysis)} We carefully handpicked 50 images that do not involve personal privacy from our smartphone albums. These images represent typical visual memory scenarios, such as public signs, document pages, and product labels. Each image is then manually annotated with:
(1) a gaze fixation indicating where users naturally look when memorizing;
(2) a structured QA pair targeting specific visual details; and
(3) a bounding box localizing the answer visually.
\bmk-core is intended as a small but high-quality reference set for validation and in-depth case studies.

\begin{figure*}
  \centering
      \includegraphics[width=\textwidth]{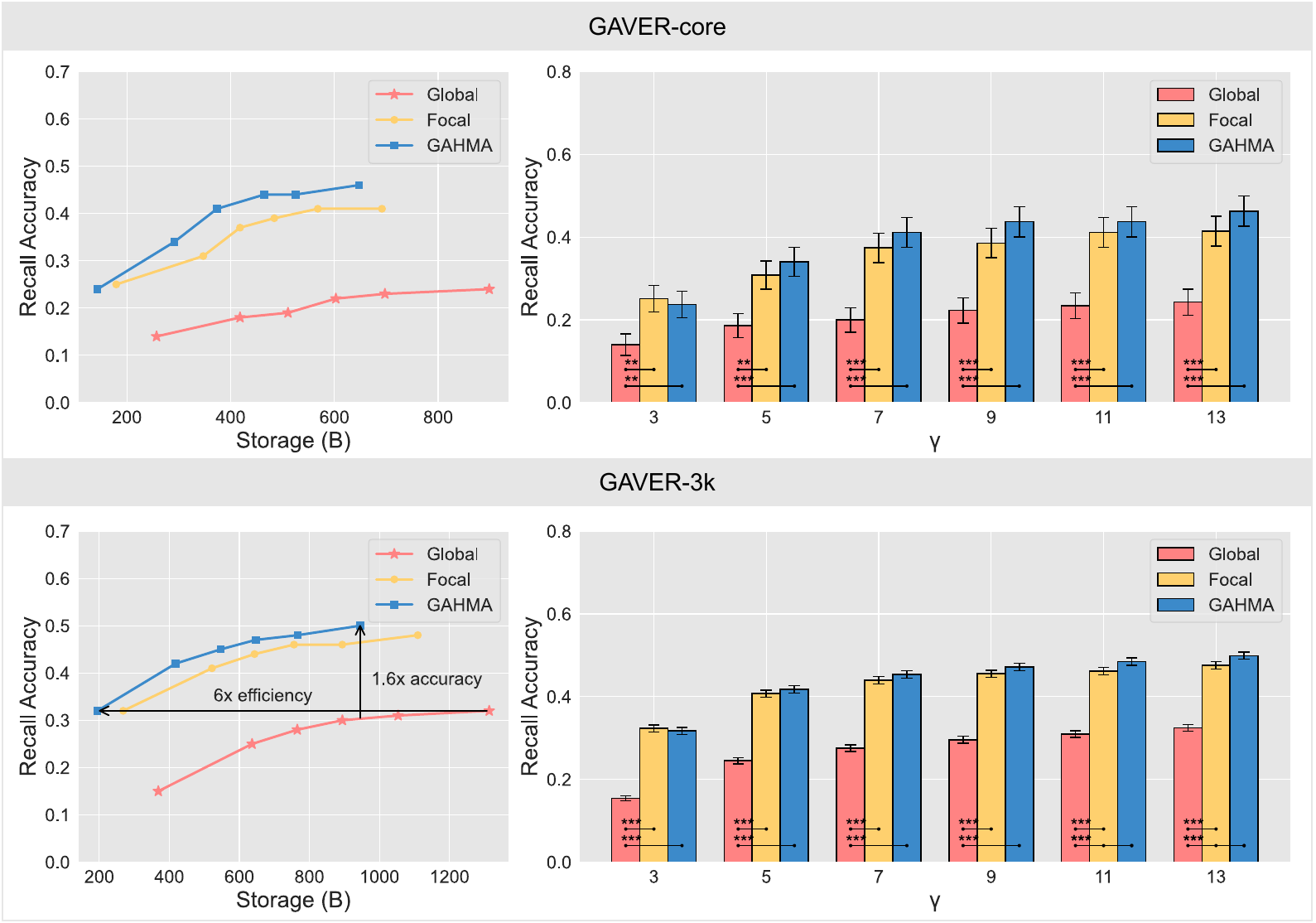}
  \caption{Results of region-specific encoding experiment. (Left) Comparison of recall accuracy and storage efficiency on \bmk-core and \bmk-3k. (Right) Significance analysis results between different encoding strategies on recall accuracy. Error bars represent standard error, with significant differences marked by ** ($p<0.01$) and *** ($p<0.001$).}
  \label{fig_exp1_1}
\end{figure*}

\textbf{\bmk-3k (web-sourced, semi-automatically annotated subset for comprehensive evaluation)} 
To further expand the scale and diversity of the dataset, we summarized a set of scene labels from \bmk-core to describe typical visual memory scenarios. Based on these labels, we searched across multiple public copyright-free platforms (e.g., Pexels, Unsplash, Pixabay) and selected 1000 images which are visually informative and semantically diverse. Complex real-world scenarios are covered such as train stations, restaurants, and office scenes, where users often wish to recall textual or symbolic content. To reduce annotation effort, we use \lm to generate visual QA pairs and simulate gaze points. The process follows five principles: (1) To simulate natural human gaze behavior, QA pairs should primarily originate from central/salient regions; (2) Questions should be grounded in visually verifiable details such as text, color, or quantity, with each answer accompanied by a bounding box specifying its visual evidence; (3) Each question must have an unambiguous answer, avoiding subjective or open-ended queries; (4) A simulated gaze point should be generated for each QA pair, located within the bounding box and typically positioned at the most visually salient subregion; and (5) In order to make full use of the collected images, at least 3 high-quality QA pairs should be generated for each image, with additional pairs generated for images containing richer semantic content.
Note that our goal is not to perfectly replicate real human gaze behavior. Instead, the \lm are used to identify and select semantically informative regions within an image. This serves as a practical way to define "user-specified points of interest" for the memory system, ensuring the dataset focuses on content likely to be deemed important for encoding, rather than generating random or uninformative locations. All generated annotations were further screened manually to remove low-quality entries (e.g., ambiguous descriptions, inaccurate boxes, or biased gaze). The final dataset contains \textbf{2936} annotated samples. 

\begin{table*}
  \caption{Numerical results in Fig.~\ref{fig_exp1_1}. Full results for all $\gamma$ are provided in Appendix \ref{full_results_of_exp1}.}
  \label{tab_exp1_1}
  \begin{tabular}{c|ccc|ccc|ccc}
    \toprule
    \cellcolor{gray!20}Dataset & \multicolumn{9}{c}{\cellcolor{gray!20}\bmk-core}  \\
    \hline
    Method & \multicolumn{3}{c|}{\cellcolor{mypink!20}Global} & \multicolumn{3}{c|}{\cellcolor{myorange!20}Focal} & \multicolumn{3}{c}{\cellcolor{myblue!20}\mtd}  \\
        \cline{2-10}
        ($\gamma$) & 3 & 7 & 13 & 3 & 7 & 13 & 3 & 7 & 13 \\
        \hline
        Recall Accuracy & 0.14 & 0.19 & 0.24 & 0.25 & 0.37 & 0.41 & 0.24 &  0.41 & 0.46  \\
        Storage(B) & 257.20 & 510.52 & 898.48 & 179.46 & 418.12 & 691.68 & 143.24 &  373.86 & 647.12  \\
    \midrule
    \cellcolor{gray!20}Dataset & \multicolumn{9}{c}{\cellcolor{gray!20}\bmk-3k}  \\
    \hline
    Method & \multicolumn{3}{c|}{\cellcolor{mypink!20}Global} & \multicolumn{3}{c|}{\cellcolor{myorange!20}Focal} & \multicolumn{3}{c}{\cellcolor{myblue!20}\mtd}  \\
        \cline{2-10}
        ($\gamma$) & 3 & 7 & 13 & 3 & 7 & 13 & 3 & 7 & 13 \\
        \hline
        Recall Accuracy & 0.15 & 0.28 & 0.32 & 0.32 & 0.44 & 0.48 & 0.32 &  0.45 & 0.50  \\
        Storage(B) & 368.66 & 765.51 & 1313.69 & 268.30 & 643.32 & 1109.66 & 194.79 &  546.74 & 944.88  \\
  \bottomrule
\end{tabular}
\end{table*}

\begin{table*}[h]
  \caption{Friedman tests ($k=3$) results for different encoding strategies on recall accuracy.}
  \label{tab_exp1_1_Friedman}
  \resizebox{\linewidth}{!}{
  \begin{tabular}{c|cccccc|cccccc}
    \toprule
    \cellcolor{gray!20}Dataset & \multicolumn{6}{c|}{\cellcolor{gray!20}\bmk-core} & \multicolumn{6}{c}{\cellcolor{gray!20}\bmk-3k} \\
    \hline
        $\gamma$ & 3 & 5 & 7 & 9 & 11 & 13 & 3 & 5 & 7 & 9 & 11 & 13 \\
        $\chi^2$ & 15.613 & 19.836 & 31.426 & 30.000 & 27.305 & 25.972 & 496.369 & 406.611 & 400.184 & 390.684 & 380.131 & 362.643 \\
        $p$ & <0.001 & <0.001 & <0.001 & <0.001 & <0.001 & <0.001 & <0.001 & <0.001 & <0.001 & <0.001 & <0.001 & <0.001 \\
  \bottomrule
\end{tabular}
}
\end{table*}

\begin{table*}[h]
  \caption{Results of post-hoc pairwise comparisons using Wilcoxon signed-rank tests for different encoding strategies on recall accuracy.}
  \label{tab_exp1_1_Wilcoxon}
  \resizebox{\linewidth}{!}{
  \begin{tabular}{c|cccccc|cccccc}
    \toprule
    \cellcolor{gray!20}Dataset & \multicolumn{6}{c|}{\cellcolor{gray!20}\bmk-core} & \multicolumn{6}{c}{\cellcolor{gray!20}\bmk-3k} \\
    \hline
        $\gamma$ & 3 & 5 & 7 & 9 & 11 & 13 & 3 & 5 & 7 & 9 & 11 & 13 \\
        $p$ (Global vs Focal) & <0.01 & <0.01 & <0.001 & <0.001 & <0.001 & <0.001 & <0.001 & <0.001 & <0.001 & <0.001 & <0.001 & <0.001 \\
        $p$ (Global vs \mtd) & <0.01 & <0.001 & <0.001 & <0.001 & <0.001 & <0.001 & <0.001 & <0.001 & <0.001 & <0.001 & <0.001 & <0.001 \\
        $p$ (Focal vs \mtd) & 0.87 & 0.23 & 0.23 & 0.20 & 0.38 & 0.22 & 0.48 & 0.17 & 0.14 & 0.13 & 0.07 & 0.09 \\
  \bottomrule
\end{tabular}
}
\end{table*}

\subsection{Methods}
The following methods are investigated in our experiments:

(1) \textbf{Global($\gamma$)} for full-image encoding. The original full image $I$ is fed into \lm to generate a global text description. This serves as a baseline for comparison.

(2) \textbf{Focal($\gamma$)} for focal-region-only encoding. The focal region $B_{f}$ is localized based on fixation point as described in Sec. \ref{sssec_gahma1}, and \lm generate a description using $I$ and $B_{f}$.

(3) \textbf{\mtdns($\gamma$)} for contextual region-based encoding. The contextual region $B_{c}$ is derived as in Eq.~\ref{eq_enclose}, and \lm generate a description using $I$, $B_{f}$, and $B_{c}$. This is the complete implementation of \mtd's encoding strategy.

In our experiments, we set $\gamma \in \{3, 5, 7, 9, 11, 13\}$ to control the maximum number of encoded sentences, defining six different detail levels, with each level building upon the previous one by adding more visual details.

We use \texttt{Qwen2.5-VL-7B-Instruct} \cite{Qwen2.5-VL} as the backbone \lm, and all experiments are conducted on NVIDIA RTX 3090 GPUs ($24$GB VRAM).

\subsection{Metrics}
In Study 1, we examine the following two objective metrics across the three methods:

(1) \textbf{Recall Accuracy} for semantic correctness of the generated answer relative to ground-truth. Each generated answer is scored by \lm on a three-point scale: 1 for complete semantic alignment with the reference (allowing minor lexical or formatting variations); 0.5 for partial semantic consistency or relevance; and 0 for semantic inconsistency or error.

(2) \textbf{Storage Usage} measuring the total byte count of stored memory entries. This metric quantifies the storage efficiency of each method.

\subsection{Evaluation of Memory Encoding via Q\&A}
\subsubsection{Region-specific encoding} \label{exp1_1}
Based on \textbf{RQ1}, we first evaluate how different region-based memory encoding strategies affect the memory enhancement. Experimental results are provided in Fig.~\ref{fig_exp1_1} and Table~\ref{tab_exp1_1}. From the results, we observe that:
(1) As $\gamma$ increases, all methods show rapid improvements in accuracy at first, followed by a gradual slowdown.
(2) Friedman tests ($k=3$) indicate significant differences between different encoding methods at each level (see Table~\ref{tab_exp1_1_Friedman} for specific values). Post-hoc pairwise comparisons using Wilcoxon signed-rank tests revealed that region-based approaches ({Focal}, \mtd) \textbf{significantly outperform} {Global} in recall accuracy across all levels (see Table~\ref{tab_exp1_1_Wilcoxon} for specific values). There were no significant differences in recall accuracy between Focal and \mtd.
(3) According to the numerical results in Table~\ref{tab_exp1_1}, \mtd generally superiors to Focal, demonstrating the supportive role of contextual region expansion.
(4) At similar accuracy levels, \mtd consumes only ~$1/6$ (or even less) of the storage used by Global, confirming its \textbf{superior storage efficiency}.

These results validate the utility of region-specific encoding strategy. They also confirm that \bmk-3k aligns well with \bmk-core in performance trends, justifying its use for large-scale evaluation.

\begin{figure*}[h]
  \centering
      \includegraphics[width=0.95\textwidth]{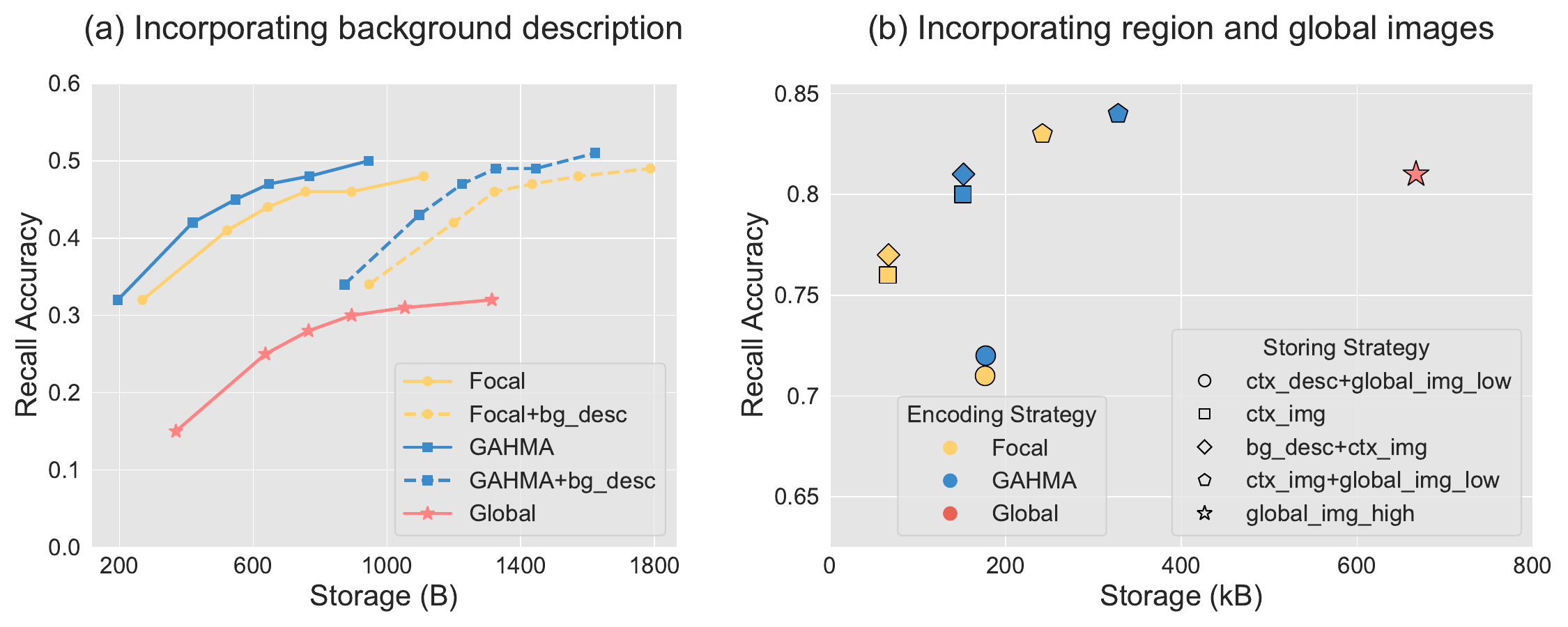}
  \caption{Recall accuracy and storage efficiency of various methods (a) with/without incorporating background description (b) under hybrid storage setting (ctx\_desc denotes the textual description generated with the corresponding encoding strategy).}
  \label{fig_exp1_2}
\end{figure*}

\begin{figure*}[h]
  \centering
      \includegraphics[width=0.95\textwidth]{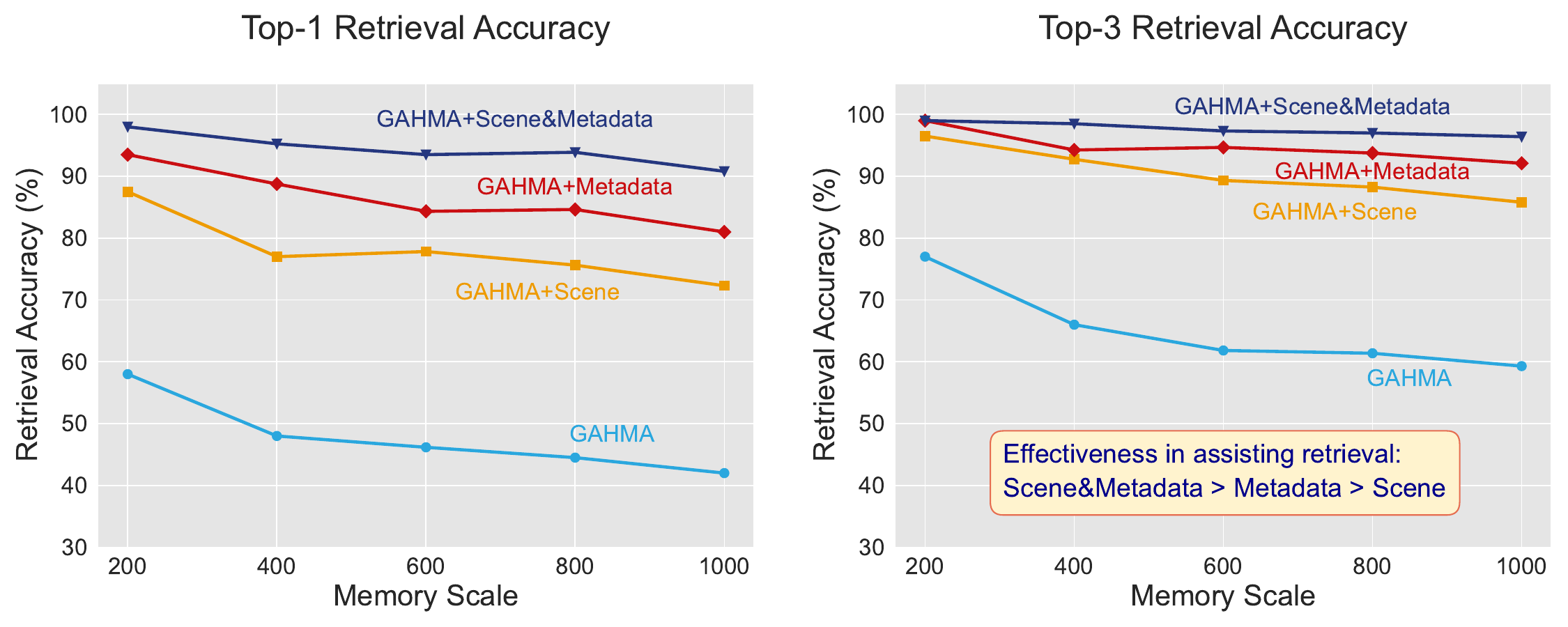}
  \caption{Retrieval accuracy of \mtd under different memory archive sizes and configurations.}
  \label{fig_exp2}
\end{figure*}

\subsubsection{Incorporating background information} \label{exp1_2}
As shown in Fig.~\ref{fig_exp1_2}(a), two region-based methods (Focal, \mtd) are evaluated with and without background summaries generated by \lm. The results indicate that: (1) The inclusion of background information has minimal impact on the overall performance trends of both methods. (2) However, the storage usage nearly doubles when background summaries are added. This indicates that our region-specific encoding strategy \textbf{already preserves task-relevant information} without demanding the background information. These findings further validate our idea of using gaze as the indicator. Detailed numerical results are provided in Appendix~\ref{full_results_of_exp1}.

According to the results of Sec.~\ref{exp1_1} and Sec.~\ref{exp1_2}, it could be concluded that the gaze-guided region-specific encoding of \mtd effectively \textbf{captures user-intended content}, addressing \textbf{RQ1}.

\subsubsection{Incorporating region and global images}
In the hybrid storage setting, methods of Focal and \mtd ($\gamma=9$, a representative configuration) are evaluated with the inclusion of high-resolution region images $I(B_f)$/$I(B_c)$ and/or low-quality global images $I_{lq}$. Performance on high-resolution global images $I$ serves as a baseline for comparison. Image patches are compressed using {JPEG} to reduce storage overhead. Results are shown in Fig.~\ref{fig_exp1_2}(b). Several key observations can be made: (1) Incorporating images \textbf{significantly boosts} recall accuracy, peaking at over $0.84$. (2) As expected, this substantially \textbf{increases memory usage} by two orders of magnitude. (3) Under the encoding strategy of \mtd, incorporating high-resolution region images can achieve comparable accuracy to using high-resolution global images, but with only less than ~$1/3$ of the storage cost. (4) Incorporating high-resolution region images and low-quality global images together yields the best performance, even surpassing the high-resolution global image baseline. (5) Considering the trade-off between accuracy and storage, \textbf{region images} offer a more favorable balance than global images. (6) Compared to using region images alone, incorporating background descriptions further improves performance. These findings suggest the influence of different storage strategies on recall accuracy and storage efficiency, addressing \textbf{RQ2}.

\begin{figure*}[t!]
  \centering
      \includegraphics[width=\textwidth]{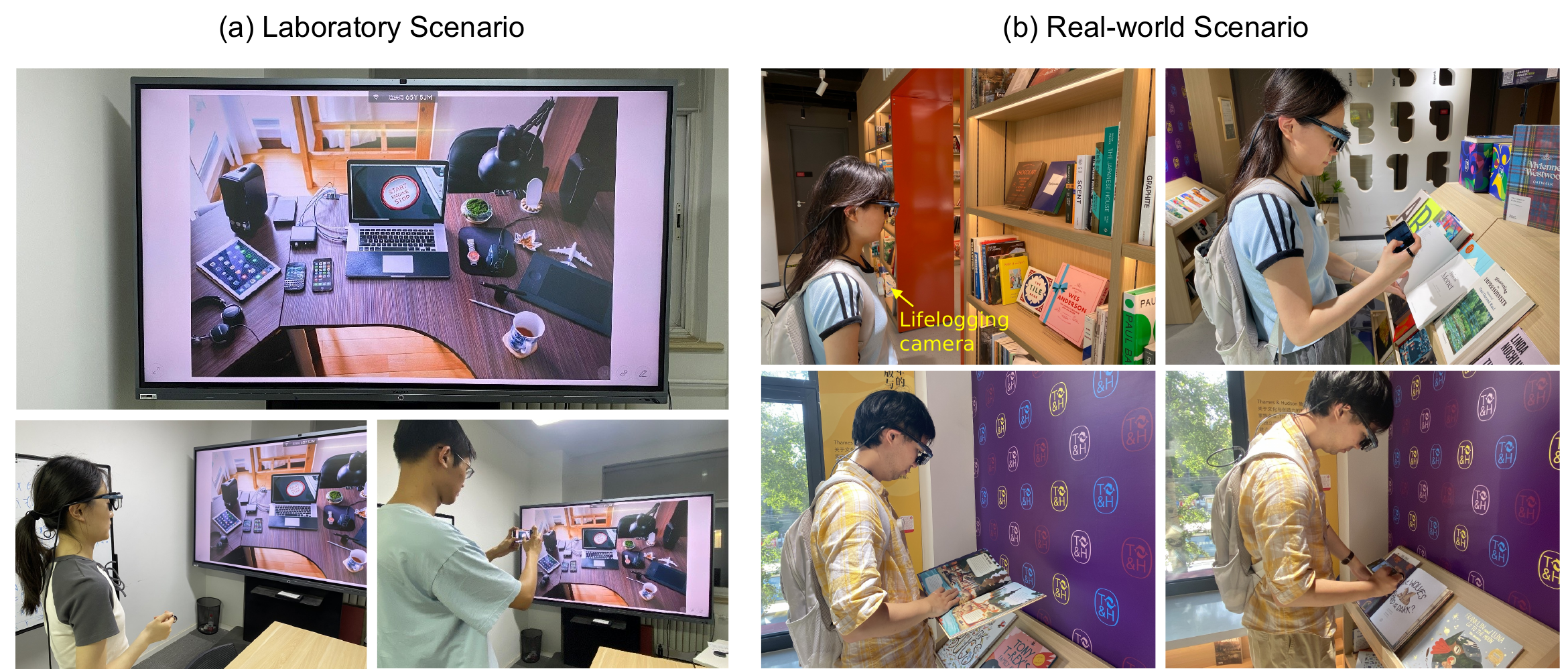}
  \caption{User evaluation in (a) laboratory scenario and (b) real-world scenario. In real-world scenario, participants wore a backpack containing a laptop, which collected all the recordings.}
  \label{fig_user_study_1}
\end{figure*}

\subsection{Evaluation of Targeted Retrieval from Memory Archives}
We evaluate the ability of \mtd ($\gamma=9$, a representative configuration) to perform targeted retrieval under a RAG-based framework. The task is to locate specific memory entries from a large set of stored memories. The vector database we use is Chroma \cite{Chromadb}, and we choose \texttt{all-MiniLM-L6-v2} \cite{embeddingModel} as the embedding model.

We make four detailed settings: (1) \mtd alone; (2) +Scene (leveraging background description for integrated retrieval); (3) +Metadata (using timestamp and GPS for pre-filtering); and (4) +Scene\&Metadata. Experiments are conducted on \bmk-3k, where we randomly sample $200$, $400$, $600$, $800$, and $1000$ images, each associated with one of its QA pair. Results are shown in Fig.~\ref{fig_exp2}, where the top-1 and top-3 retrieval accuracies of \mtd alone drop continuously as memory size increases, falling about $40\%$ and $60\%$ respectively at $1000$ entries. In contrast, when assisted by scene descriptions and metadata, the top-3 retrieval accuracy maintains $96.4\%$ even with $1000$ stored entries. This demonstrates that our method can effective retrieval from \textbf{large-scale memory archives}, which answers \textbf{RQ3}.

\section{Study 2: Usability Study for \ga}

\subsection{Research Objectives}
Study 1 demonstrated the technical effectiveness of \mtd in terms of memory encoding and retrieval. To further assess the practical usability and user experience of \ga, we conduct a user study with the following research question:

\textbf{RQ5: How do users perceive the usability of \ga compared to existing visual memory augmentation approaches, particularly in terms of user effort, unobtrusiveness, and overall preference?} This question mainly evaluates whether \ga satisfies the second fundamental requirement of effective memory augmentation systems.

\subsection{Participants}
We recruited 16 participants (4 female, 12 male) aged between 22
and 30 years ($M$ = 24.56, $SD$ = 2.09), and 13 of them wore glasses. Participants rated their experience with smart wearables usage ($M$ = 3.82, $SD$ = 0.81), gaze interaction ($M$ = 3.56, $SD$ = 0.93) and ring interaction ($M$ = 2.31, $SD$ = 1.04) on a scale from 1 (no experience) to 5 (expert). The study received ethics approval from the university ethics review board, and participants gave written consent to take part in the study.

\subsection{Apparatus}
To assess the usability and subjective user experience of \ga, we implement it with \mtd on a pair of smart glasses\footnote{https://www.7invensun.com/asee\_glasses} equipped with an eye-tracking module and an egocentric camera (1280{\scriptsize $\times$}960 px), along with a Bluetooth-enabled ring. The smart glasses and the ring are connected to a laptop (Lenovo IdeaPad Pro 5 16", i5-13500H) via TCP and Bluetooth, respectively.

For comparison, we implement another two methods: (1) a \textbf{Phone-based} baseline on an Android device with a 6.36" screen and a camera of 4096{\scriptsize $\times$}3072 px resolution, where users capture scenes manually (zooming, framing, and triggering) using smartphone cameras, and (2) a \textbf{Lifelogging} baseline with a thumb-sized camera\footnote{https://item.jd.com/10165873142487.html} (1920{\scriptsize $\times$}1080 px) mounted on the chest, which continuously records the user's daily activities. Both baseline methods use the same \lm for encoding, and Lifelogging method extracts four frames per second from the recorded video, following the configuration in \cite{shen2024encode}.

\subsection{Procedure}
The user study consists of two scenarios: a controlled laboratory scenario and a real-world bookstore scenario.

\subsubsection{Laboratory Scenario}

The experimental setup was in a room with a large screen. The task consisted of 8 trails per participant. In each trail, participants listened to a short audio clip while simultaneously memorizing a scene displayed on the screen using \ga or Phone-based method (4 trails per method). Immediately after the audio clip ended, participants were asked to answer a question about the just-played audio clip.

Scenes were sampled from \bmk-3k. Participants were given one question per scene, and asked to find answers and record relevant content using either \ga or a smartphone.

Before the experiment, participants received a comprehensive explanation of the study's purpose and procedure through a PowerPoint presentation. Subsequently, a brief training session was conducted, during which participants were introduced to both recording modalities using four practice images to ensure proficiency and minimize learning efforts during the formal trials.

In the main study, each trail involved the presentation of an image on the large screen, accompanied by a question designed to guide participants' attention toward specific visual target. In the Phone-based condition, participants were instructed to first identify the visual element that would serve as the answer to the given question within five seconds. An audio clip (approximately 20 seconds) was then played. Two seconds after audio onset, a "start capturing" cue appeared on the screen, indicating that participants should begin recording. Specifically, participants retrieved the phone from an adjacent table, launched the camera application, adjusted the framing and zoom according to their personal preferences, and finally triggered the shutter. This procedure was intended to emulate real-world scenarios where users may encounter and wish to record information opportunistically. After the audio clip concluded, participants answered a question related to the audio content.

In the \ga condition, participants followed a similar process to identify the target visual element, listen to an audio clip while capturing, and answer a question. After the "start capturing" cue, participants fixated their gaze on the target object and initiated recording by performing a double-tap gesture on the Bluetooth ring. 

To mitigate potential order and learning effects, we employed a counterbalanced within-subjects experimental design. Participants were grouped in pairs, and each pair was assigned two sets of four images. The assignment of recording methods to image sets was reversed across participants within each pair. Furthermore, the test order of the two recording methods was alternated across participant pairs, and the presentation order of the audio clips was systematically varied using a Latin square design.

\subsubsection{Real-world Scenario}
We invited 3 participants (2 female, 1 male) from our in-lab experiment to simultaneously wear the smart glasses and lifelogging camera while freely exploring a bookstore for 10 minutes. During the exploration, they were asked to record any information they found interesting or wished to remember using all three methods (Lifelogging baseline automatically recorded the entire process without explicit interaction). This scenario aimed to evaluate the perceived unobtrusiveness and practicality of \ga in real-world situations.

\subsection{Results}


\subsection{Evaluation Metrics}
\subsubsection{Objective Metrics}
In the laboratory scenario, we examine three objective metrics across \ga and Phone-based visual memory augmentation methods:

(1) \textbf{Recording Time.} This metric quantifies the time taken to complete the recording task, from the moment the "start capturing" cue appears to when the image is finally logged.

(2) \textbf{Visual recall Accuracy.} This metric serves as an indicator of the quality and informativeness of the captured visual content. To control for variations in image quality, all user-captured images (from both devices) were aligned to their corresponding source images using a perspective transformation. These aligned images were subsequently processed using \mtd or Global encoding strategy. The encoded outputs were then used to answer the associated questions, and the accuracy of these responses were measured in the same way as in Study 1.

(3) \textbf{Audio QA Accuracy.} This metric measures participants' correctness on audio-related questions (binary: 1 = correct, 0 = incorrect). It is used to evaluate the degree of interference introduced by each method on the ongoing task.

In the real-world scenario, we compare the \textbf{Visual Recall Accuracy} of \ga and Lifelogging baseline, with smartphone-captured images serving as the reference for correct answers. In addition, we evaluate the \textbf{Storage Usage} of the two methods. 

\subsubsection{Subjective Metrics}
In both scenarios, we collected participants' feedback through a questionnaire regarding their experiences with the three memory augmentation systems. The questionnaire assessed aspects such as \textbf{physical effort}, \textbf{disruption}, \textbf{social obtrusiveness} and \textbf{overall preference} on a 5-point Likert scale (1 = strongly disagree, 5 = strongly agree). Additionally, open-ended questions were included to collect feedback on the overall experience in the study and suggestions for improvement.

\begin{figure}[t!]
  \centering
      \includegraphics[width=\linewidth]{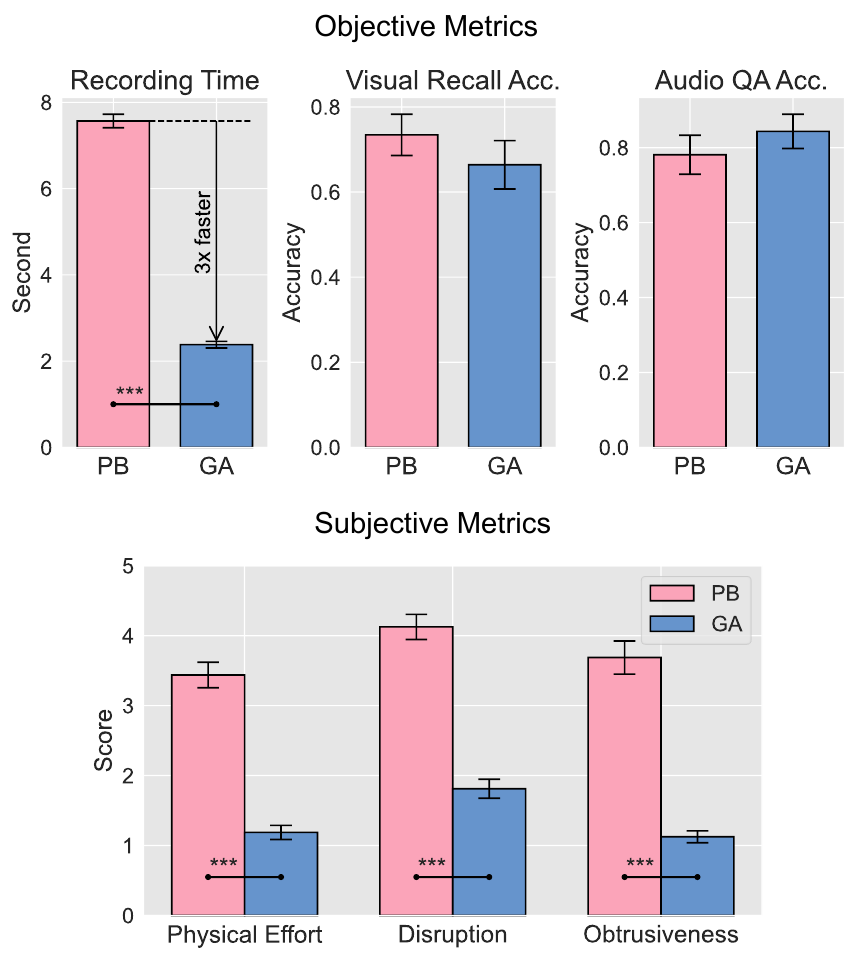}
  \caption{Objective and subjective results of in-lab user study (PB = Phone-based, GA = \ga). Error bars denote the standard deviation, with significant differences indicated by *** ($p$ < 0.001).}
  \label{fig_user_study_1_res}
\end{figure}

\subsubsection{Laboratory scenario} 
The objective results of in-lab user study are shown in the upper panel of Fig.~\ref{fig_user_study_1_res}. A paired-samples t-test revealed a significant differences in \textbf{recording time} between \ga and Phone-based method ($t(63) = 29.608$, $p < 0.001$, Cohen's $d = 3.701$). The average recording time for \ga (2.38 s) was significantly shorter than that of Phone-based (7.57 s). This indicates that \ga enables users to capture visual memories more quickly and efficiently. As for \textbf{visual recall accuracy}, there were no significant differences between the two methods ($t(63) = 0.98$, $p = 0.33$, Cohen's $d = 0.122$), with \ga achieving an average accuracy of 0.66 and Phone-based achieving 0.73. This suggests that \ga can capture visual information of comparable quality to traditional Phone-based methods. Regarding \textbf{audio QA accuracy}, there were no significant differences between the two methods ($t(63) = 1.16$, $p = 0.25$, Cohen's $d = 0.145$), with \ga achieving an average accuracy of 0.84 and Phone-based achieving 0.78.

The subjective results are shown in the lower panel of Fig.~\ref{fig_user_study_1_res}. A paired-samples t-test indicated significant differences in perceived \textbf{physical effort} between the two methods ($t(15) = 13.17$, $p < 0.001$, Cohen's $d = 3.29$). Participants rated \ga (M = 0.19, SD = 0.39) as requiring significantly less effort than Phone-based (M = 3.44, SD = 0.70). This suggests that \ga is easier to use and requires less physical exertion.
Though objective audio QA accuracy showed no significant differences between the two methods, a paired-samples t-test on the subjective scores of \textbf{disruption} revealed significant differences in user perception ($t(15) = 15.36$, $p < 0.001$, Cohen's $d = 3.841$). Participants rated \ga (M = 1.81, SD = 0.53) as causing significantly less disruption than Phone-based (M = 4.13, SD = 0.70). This indicates that \ga is less intrusive and allows users to maintain better focus on their primary tasks. For \textbf{social obtrusiveness}, a paired-samples t-test revealed significant differences between the two methods ($t(15) = 12.59$, $p < 0.001$, Cohen's $d = 3.148$). Participants rated \ga (M = 1.13, SD = 0.33) as significantly less obtrusive than Phone-based (M = 3.69, SD = 0.92). This suggests that \ga is less noticeable to others and allows users to record memories more privately.
All 16 participants expressed a \textbf{preference for \ga over Phone-based}, indicating a unanimous favorability towards the gaze-aware system.

\begin{figure}[t!]
  \centering
      \includegraphics[width=\linewidth]{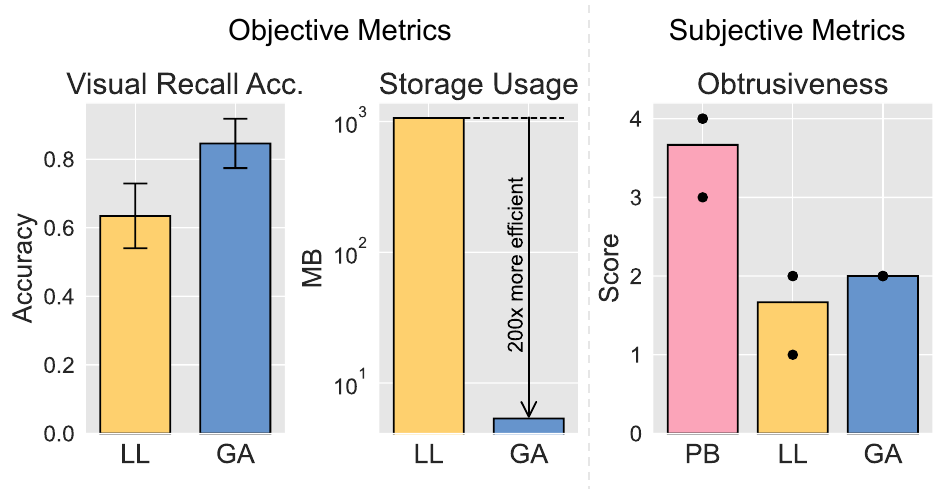}
  \caption{Objective and subjective results of in-situ user study (PB = Phone-based, LL = Lifelogging, GA = \ga). Error bars denote the standard deviation.}
  \label{fig_user_study_2_res}
\end{figure}

\subsubsection{Real-world scenario}
\begin{figure*}[htbp]
  \centering
      \includegraphics[width=0.99\textwidth]{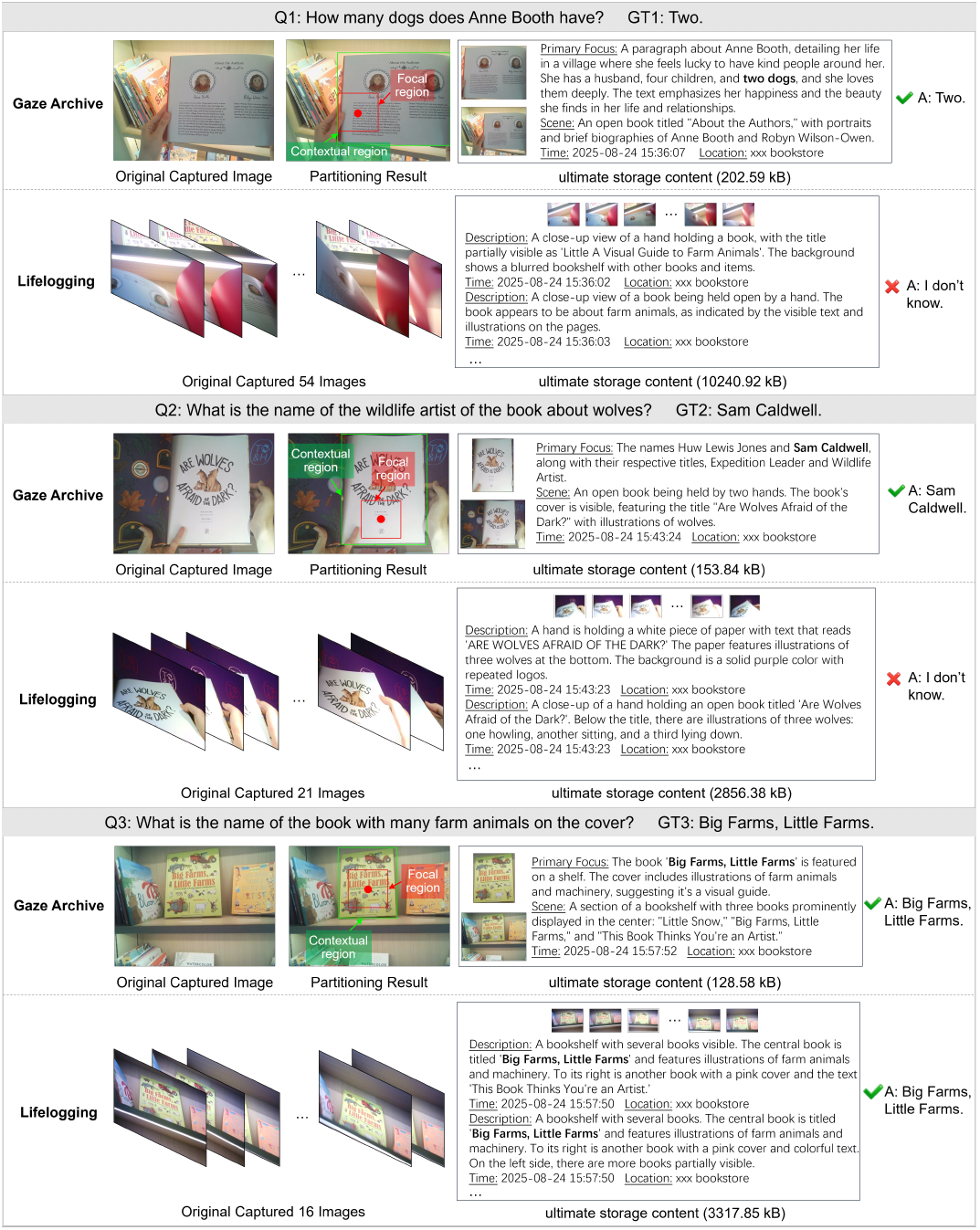}
  \caption{Three representative examples of user-captured images and corresponding QA performance with \ga and Lifelogging in real-world scenario. Note that the storage usage only includes the statistics of content related to the question in Lifelogging method.}
  \label{fig_user_study_case}
\end{figure*}

The objective and subjective results of in-situ user study are shown in Fig.~\ref{fig_user_study_2_res}. Participants have made a total of 26 records. Three representative examples of user-captured images and corresponding QA performance with \ga and Lifelogging are shown in Fig.~\ref{fig_user_study_case}. The average \textbf{visual recall accuracy} for \ga was 0.85 (SD = 0.36), while for Lifelogging it was 0.63 (SD = 0.47). This was because \ga captures user experiences from participants' first-person perspective, whereas Lifelogging camera, constrained by its fixed chest-mounted viewpoint, may not always succeed in comprehensively capturing the intended content, as shown in the first two examples of Fig.~\ref{fig_user_study_case}. In terms of \textbf{storage usage}, Gaze Archive occupies a total of 5.36 MB, while lifelogging occupies a total of 1063.65 MB. This indicates that \ga is significantly more storage-efficient, as it only captures user-intended content rather than continuous recording. As for subjective ratings, participants rated \ga (M = 2) and Lifelogging (M = 1.67) as less obtrusive than Phone-based method (M = 3.67) in terms of \textbf{social obtrusiveness}. All participants expressed a \textbf{preference for \ga over Lifelogging and Phone-based}, indicating its promising utility in real-world scenarios.

The results in both scenarios demonstrate the interaction advantages of \ga, addressing \textbf{RQ5}.

\section{Discussion}
In this paper, we propose a novel visual memory augmentation paradigm designed to meet the dual requirements of intent-precise memory capture and unobtrusive logging with low effort. Based on the results of Study 1 and Study 2, we further discuss how \ga performs in these two aspects, along with users' qualitative feedback and future design implications.


\subsection{Effectiveness of \ga in Intent-Precise Memory Capture}
In study 1, our results demonstrate that \mtd's gaze-guided region-specific encoding strategy effectively captures user-intended content. Specifically, compared to full-image encoding, \mtd significantly improves recall accuracy while greatly reducing storage usage. This indicates that prioritizing gaze-indicated regions allows the system to capture the most relevant information for memory tasks. Incorporating region and global images further enhances recall accuracy, and is more preferred by users than text-only storage. In the real-world scenario, \ga also outperforms Lifelogging baseline in recall accuracy, and achieves comparable performance to Phone-based capture, which is regarded as the most direct way to capture user-intended content. These findings confirm that \mtd can effectively capture and store visual memories aligned with user intent, addressing the first fundamental requirement of memory augmentation systems.


\subsection{Effectiveness of \ga in Effortless and Unobtrusive Interaction}
Study 2 validates \ga's advantage in providing an effortless and unobtrusive memory logging experience. The results from the controlled laboratory scenario indicate that \ga enables significantly faster and more effortless operation compared to traditional Phone-based methods. Many participants also appreciated the natural and intuitive characteristics of gaze-based interaction. As {\itshape P9} noted, "using the smart glasses felt much more natural and less cumbersome than pulling out my phone every time I wanted to capture something". Additionally, \ga allows users to maintain focus on concurrent tasks with minimal cognitive load, as evidenced by the higher audio QA accuracy and significant lower disruption score. In the real-world scenario, \ga also exhibits promising utility in perceived social unobtrusiveness. In both scenarios, \ga received unanimous preference over existing methods. These findings confirm that \ga effectively meets the second fundamental requirement of effective memory augmentation systems.


\subsection{Other Qualitative Feedback and Design Implications} \label{ssec_qualitative_feedback}
At the end of the user study, participants provided valuable qualitative feedback on their experiences with different memory augmentation methods and offered suggestions for future improvements. From the open-ended survey, we observed that participants emphasized the value of \ga, highlighting various applicable scenarios such as lectures, meetings, shopping and exploring new places. In terms of logging devices, {\itshape P1} expressed preference for smart glasses over traditional chest-mounted lifelogging cameras, stating that the former "allows her to capture exactly what she's looking at, rather than just a general view of her surroundings". This also Illustrates the strength of \ga in capturing intent-aligned visual memories. Despite the positive feedback, some participants also highlighted the wearability of smart glasses for prolonged use since it's heavier than ordinary glasses {\itshape (P3)}, particularly for those who do not regularly wear glasses {\itshape{(P8)}}. This suggests that future designs could further improve the comfort and ergonomics of the wearable hardware. Additionally, while appreciating the convenience and unobtrusiveness of \ga in public settings, some participants expressed concerns about the privacy issues of wearing a camera-equipped eyewear in social contexts. For example, {\itshape P7} noted that "compared with using a smartphone, taking photos with smart glasses and a ring is much more discreet, which may lead to misuse", while {\itshape P14} and {\itshape P15} expressed worries about "discomfort with others wearing the system and recording themselves without consent". These worries ought to be fully taken into account in future designs.

\section{Limitations, Future Work and Conclusion}
\subsection{Limitations and Future Work}
While we have demonstrated the advantages and effectiveness of \ga in both quantitative evaluations and user studies, there remain several limitations and avenues for future research.
\subsubsection{Study Design and Population}
Firstly, our study was primarily conducted with university students, who may be more accustomed to such technology as wearable assistants. Future studies should include a more diverse demographic to ensure broader applicability. Secondly, the controlled laboratory scenario may not fully capture the complexities and unpredictability of real-world environments. Although we included a real-world bookstore scenario, it was limited to a short duration and a small number of participants. Future research should explore large-scale, longer-term deployments across various real-life scenarios to better understand system performance and user experience over time, as well as the long-term effects on human memory and behavior.

\subsubsection{System Limitations}
Firstly, the performance of \mtd is inherently tied to the capabilities of the underlying \lm. As a prototype, our implementation uses \texttt{Qwen2.5-VL-7B-Instruct} \cite{Qwen2.5-VL} as the backbone \lm. While this model demonstrates strong performance in our evaluation, current advancements in AI may soon surpass the capabilities of the system implemented in this work. Secondly, the current system depends on wireless access to \lm, which may not be feasible in various real-world scenarios. Deploying \lm on-device remains a promising direction. Thirdly, \lm are known to suffer from hallucinations, occasionally producing fabricated or misleading information. Such issues may compromise the reliability of memory recall and mislead users. Addressing this challenge calls for the joint efforts of both the AI and HCI communities to develop robust solutions that ensure the accuracy and trustworthiness of generated content. Moreover, our current implementation only supports static visual memory. Future work could explore incorporating dynamic visual content and additional modalities such as auditory information, to create a more comprehensive wearable memory augmentation system. Finally, as mentioned in Sec.~\ref{ssec_qualitative_feedback}, practical use in daily life requires parallel advancement in smart glasses technology and privacy protection mechanisms such as automatic redaction of sensitive information should be considered.



\subsection{Conclusion}
In this paper, we propose \ga, a novel visual memory augmentation paradigm through active logging on smart glasses, aiming to achieve both intent-precise memory capture and effortless-and-unobtrusive interaction. Users simply need to naturally gaze at the desired target and perform a double-tap gesture on a ring to complete memory logging. Building upon this new paradigm, we present \mtd, a practical technical framework that leverages human gaze to enable intent-precise memory encoding and retrieval. Quantitative evaluations on the newly introduced \bmk benchmark dataset demonstrate \mtd's superior performance compared to non-gaze baselines. Comprehensive user studies conducted in both laboratory and real-world scenarios validate the perceived effortlessness and unobtrusiveness of \ga, as well as its overall user preference over traditional Phone-based and Lifelogging methods. We also identified key considerations for future designs of memory augmentation systems, including deploying large models on-device, extending multimodal memory, and balancing unobtrusiveness and privacy.

\bibliographystyle{ACM-Reference-Format}
\bibliography{ref}

\appendix
\section{Complete Quantitative Experiment Results}

\subsection{Results on Memory Encoding via Q\&A} \label{full_results_of_exp1}
Here we provide full numerical results in Fig.~\ref{fig_exp1_1} and \ref{fig_exp1_2}.

\begin{table}[H]
  \caption{Full numerical results on \bmk-core in Fig.~\ref{fig_exp1_1}.}
  \label{tab_core_complete}
  \centering
  \resizebox{\linewidth}{!}{
      \begin{tabular}{c|cccccc}
        \toprule
        Global($\gamma$) & 3 & 5 & 7 & 9 & 11 & 13 \\
        Recall Acc. & 0.14 & 018 & 0.19 & 0.22 & 0.23 & 0.24 \\
        Stor.(B) & 257.20 & 418.14 & 510.52 & 602.82 & 697.38 & 898.48 \\
        \midrule
        Focal($\gamma$) & 3 & 5 & 7 & 9 & 11 & 13 \\
        Recall Acc. & 0.25 & 0.31 & 0.37 & 0.39 & 0.41 & 0.41 \\
        Stor.(B) & 179.46 & 347.28 & 418.12 & 484.12 & 568.04 & 691.68 \\
        \midrule
        \mtdns($\gamma$) & 3 & 5 & 7 & 9 & 11 & 13 \\
        Recall Acc. & 0.24 & 0.34 & 0.41 & 0.44 & 0.44 & 0.46 \\
        Stor.(B) & 143.24 & 291.56 & 373.86 & 464.90 & 525.24 & 647.12 \\
        \bottomrule
      \end{tabular}
  }
\end{table}

\begin{table}[H]
  \caption{Full numerical results on \bmk-3k in Fig.~\ref{fig_exp1_1}.}
  \centering
  \resizebox{\linewidth}{!}{
      \begin{tabular}{c|cccccc}
        \toprule
        Global($\gamma$) & 3 & 5 & 7 & 9 & 11 & 13 \\
        Recall Acc. & 0.15 & 0.25 & 0.28 & 0.30 & 0.31 & 0.32 \\
        Stor.(B) & 368.66 & 636.38 & 765.51 & 894.28 & 1053.53 & 1313.69 \\
        \midrule
        Focal($\gamma$) & 3 & 5 & 7 & 9 & 11 & 13 \\
        Recall Acc. & 0.32 & 0.41 & 0.44 & 0.46 & 0.46 & 0.48 \\
        Stor.(B) & 268.30 & 521.88 & 643.32 & 756.13 & 894.02 & 1109.66 \\
        \midrule
        \mtdns($\gamma$) & 3 & 5 & 7 & 9 & 11 & 13 \\
        Recall Acc. & 0.32 & 0.42 & 0.45 & 0.47 & 0.48 & 0.50 \\
        Stor.(B) & 194.79 & 418.26 & 546.74 & 646.92 & 767.35 & 944.88 \\
        \bottomrule
      \end{tabular}
  }
\end{table}

\begin{table}[H]
  \caption{Full numerical results in Fig.~\ref{fig_exp1_2}(a).}
  \label{tab_bg_and_img_patch_complete_focal}
  \centering
  \resizebox{\linewidth}{!}{
      \begin{tabular}{c|cccccc}
        \toprule
        Focal($\gamma$)+bg\_desc & 3 & 5 & 7 & 9 & 11 & 13  \\
        Recall Acc. & 0.34 & 0.42 & 0.46 & 0.47 & 0.48 & 0.49 \\
        Stor.(B) & 946.33 & 1199.92 & 1321.36 & 1434.20 & 1572.09 & 1787.74 \\
        \midrule
        \mtdns($\gamma$)+bg\_desc & 3 & 5 & 7 & 9 & 11 & 13  \\
        Recall Acc. & 0.34 & 0.43 & 0.47 & 0.49 & 0.49 & 0.51 \\
        Stor.(B) & 872.68 & 1096.16 & 1224.64 & 1324.83 & 1445.26 & 1622.79 \\
        \bottomrule
      \end{tabular}
  }
\end{table}

\begin{table}[H]
  \caption{Full numerical results in Fig.~\ref{fig_exp1_2}(b).}
  \label{tab_bg_and_img_patch_complete_focal}
  \centering
  \resizebox{\linewidth}{!}{
      \begin{tabular}{c|cc|cc}
        \toprule
        \multirow{2}{*}{Strategy} &  \multicolumn{2}{c|}{Focal(9)} & \multicolumn{2}{c}{\mtdns(9)} \\
        ~ & Recall Acc. & Stor.(kB) & Recall Acc. & Stor.(kB) \\
        \midrule
        ctx\_desc+global\_img\_low & 0.71 & 176.54 & 0.72 & 177.26 \\
        ctx\_img & 0.76 & 66.02 & 0.80 & 151.28 \\
        bg\_desc+ctx\_img & 0.77 & 66.68 & 0.81 & 151.95 \\
        ctx\_img+global\_img\_low & 0.83 & 241.82 & 0.84 & 327.92 \\
        \bottomrule
      \end{tabular}
  }
\end{table}


\subsection{Results on Targeted Retrieval from Memory Archives}
Here we provide full numerical results in Fig.~\ref{fig_exp2}.

\begin{table}[H]
  \caption{Top-1 Retrieval Accuracy of Different Strategies on Memory Archives Retrieval Task.}
  \label{tab_archives}
  \centering
  \resizebox{\linewidth}{!}{
      \begin{tabular}{c|ccccc}
        \toprule
        \multirow{2}{*}{Method} &  \multicolumn{5}{c}{Memory Scale} \\
        ~ & 200 & 400 & 600 & 800 & 1000 \\
        \midrule
        \mtdns(9) & 58.00\% & 48.00\% & 46.17\% & 44.50\% & 42.00\% \\
        \mtdns(9)+Scene & 87.50\% & 77.00\% & 77.83\% & 75.63\% & 72.30\% \\
        \mtdns(9)+Metadata & 93.50\% & 88.75\% & 84.33\% & 84.63\% & 81.00\% \\
        \mtdns(9)+Scene\&Metadata & 98.00\% & 95.25\% & 93.50\% & 93.88\% & 90.80\% \\
        \bottomrule
      \end{tabular}
  }
\end{table}

\begin{figure*}
  \centering
  \includegraphics[width=0.93\textwidth]{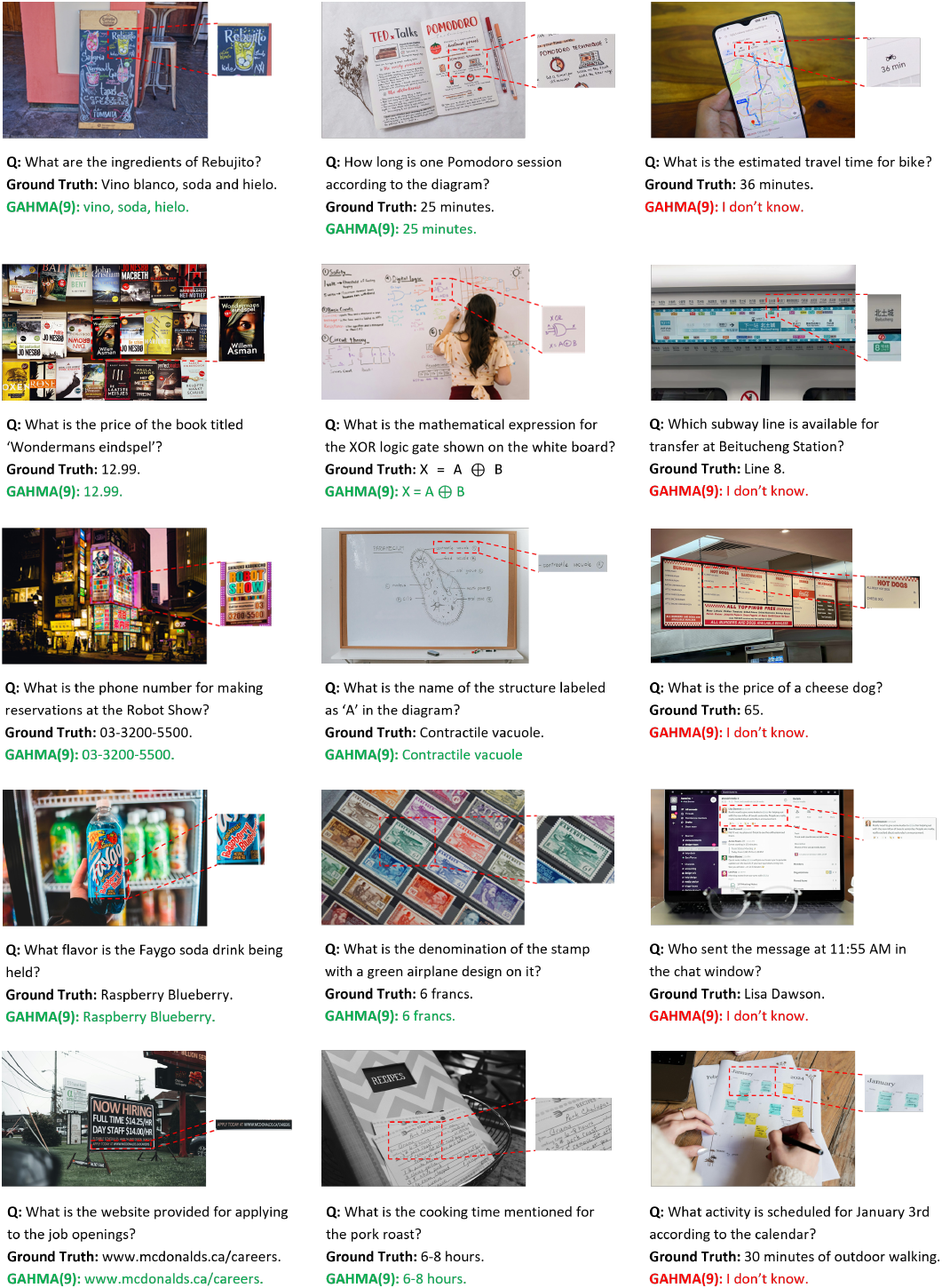}
  \caption{Examples of QA pairs from \bmk-3k, along with the corresponding performance of \mtdns(9).}
  \label{fig_qa_examples}
\end{figure*}

\begin{table}[H]
  \caption{Top-3 Retrieval Accuracy of Different Strategies on Memory Archives Retrieval Task.}
  \label{tab_archives}
  \centering
  \resizebox{\linewidth}{!}{
      \begin{tabular}{c|ccccc}
        \toprule
        \multirow{2}{*}{Method} &  \multicolumn{5}{c}{Memory Scale} \\
        ~ & 200 & 400 & 600 & 800 & 1000 \\
        \midrule
        \mtdns(9) & 77.00\% & 66.00\% & 61.83\% & 61.38\% & 59.30\% \\
        \mtdns(9)+Scene & 96.50\% & 92.75\% & 89.33\% & 88.25\% & 85.80\% \\
        \mtdns(9)+Metadata & 99.00\% & 94.25\% & 94.67\% & 93.75\% & 92.10\% \\
        \mtdns(9)+Scene\&Metadata & 99.00\% & 98.50\% & 97.33\% & 97.00\% & 96.40\% \\
        \bottomrule
      \end{tabular}
  }
\end{table}

\section{Dataset Example}  \label{sec_case_study}

In Fig.~\ref{fig_qa_examples}, we show some examples of the QA pairs from our \bmk dataset, as well as the results of the representative method \mtd(9) on them, including both successful and failed cases. 

\end{document}